\documentclass{IEEEtran4PSCC}
\ifCLASSINFOpdf
   \usepackage[pdftex]{graphicx}
\else
   \usepackage[dvips]{graphicx}
\fi
\usepackage[cmex10]{amsmath}
\usepackage{amsmath,amsfonts}
\usepackage{algorithmic}
\usepackage{algorithm}
\usepackage{array}
\usepackage[caption=false,font=normalsize,labelfont=sf,textfont=sf]{subfig}
\usepackage{textcomp}
\usepackage{stfloats}
\usepackage{url}
\usepackage{verbatim}
\usepackage{cite}
\usepackage{xcolor}
\usepackage{csquotes} 
\usepackage{lineno}
\usepackage{graphicx}
\usepackage{caption}
\usepackage{subfig}
\usepackage{adjustbox}
\usepackage{hyperref}
\usepackage{bm}
\usepackage{multirow}
\usepackage{tabularx} 
\usepackage{tikz}
\usepackage{tabularx, colortbl} 
\usepackage{tabularx, booktabs} 

\definecolor{lightgray}{gray}{0.9}
\usetikzlibrary{shapes.geometric, arrows}

\tikzstyle{startstop} = [ellipse, minimum width=2cm, minimum height=1cm,text centered, draw=black]
\tikzstyle{process} = [rectangle, minimum width=1.5cm, minimum height=1cm, text centered, draw=black]
\tikzstyle{decision} = [diamond, aspect=2, text centered, draw=black]
\tikzstyle{data} = [trapezium, trapezium left angle=70, trapezium right angle=110,
                    minimum width=1cm, minimum height=1cm, text centered, draw=black]
\tikzstyle{arrow} = [thick,->,>=stealth]

\makeatletter
\let\old@ps@headings\ps@headings
\let\old@ps@IEEEtitlepagestyle\ps@IEEEtitlepagestyle
\def\psccfooter#1{%
    \def\ps@headings{%
        \old@ps@headings%
        \def\@oddfoot{\strut\hfill#1\hfill\strut}%
        \def\@evenfoot{\strut\hfill#1\hfill\strut}%
    }%
    \def\ps@IEEEtitlepagestyle{%
        \old@ps@IEEEtitlepagestyle%
        \def\@oddfoot{\strut\hfill#1\hfill\strut}%
        \def\@evenfoot{\strut\hfill#1\hfill\strut}%
    }%
    \ps@headings%
}
\makeatother

\psccfooter{%
        \parbox{\textwidth}{\hrulefill \\ \small{24th Power Systems Computation Conference} \hfill \begin{minipage}{0.2\textwidth}\centering \vspace*{4pt} \includegraphics[scale=0.06]{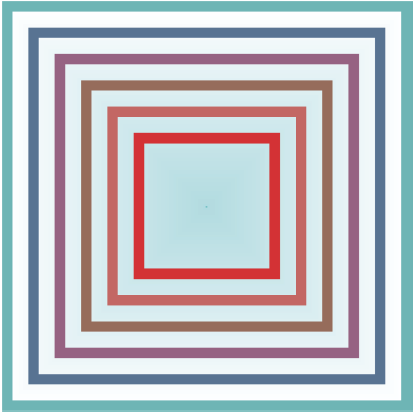}\\\small{PSCC 2026} \end{minipage} \hfill \small{Limassol, Cyprus --- June 8-12, 2026}}%
}

\begin{document}
%
\title{Equivalent Circuit Modeling of Grid-Forming Inverters in (Sub)-Transient Time-Frame}



\author{\IEEEauthorblockN{Ambuj~Gupta\IEEEauthorrefmark{1},
Balarko~Chaudhuri\IEEEauthorrefmark{1},
Mark~O'Malley\IEEEauthorrefmark{1}}
\IEEEauthorblockA{\IEEEauthorrefmark{1}Department of Electrical and Electronic Engineering\\
Imperial College London, United Kingdom\\
\{a.gupta23, b.chaudhuri, m.omalley\}@imperial.ac.uk}
}

\maketitle

\begin{abstract}
The widely accepted definition of grid-forming (GFM) inverter states that it should behave as a (nearly) constant voltage source behind an impedance by maintaining a (nearly) constant internal voltage phasor in the sub-transient to transient time frame. Some system operators further mandate permissible ranges for this effective impedance. However, these specifications do not clearly define the location of the internal voltage source, and no systematic method exists to quantify its effective impedance for a black-box GFM model. To address this, we first compare the transient responses of an ideal voltage source and a GFM to show that an idealistic GFM maintains a (nearly) constant voltage across the filter capacitor, rather than at the inverter switches. Then we propose a systematic method to quantify the effective impedance of a GFM from its black-box model using frequency-domain admittance plots. \textcolor{black}{Using standard PSCAD GFM models developed by NLR (formerly NREL), we demonstrate that the GFM's equivalent impedance model captures the sub-transient response and static voltage stability limit accurately. Further, replacing the GFM with the proposed equivalent circuit model in the modified IEEE-39 bus system is shown to reproduce the small-signal stability characteristics with reasonable accuracy}.
\end{abstract}%
%
\thanksto{\noindent This work was funded by a Leverhulme International Professorship, grant reference [LIP-2020-002] and by the Engineering and Physical Sciences Research Council [EP/Y025946/1].\\
Submitted to the 24th Power Systems Computation Conference (PSCC 2026).}%
\section{Introduction}

The growing use of inverter-based resources (IBRs) in power systems raises stability concerns, with Grid-Forming (GFM) IBRs seen as a promising solution over the commonly used Grid-Following IBRs. Addressing the need for clear definitions and performance criteria, regulators, system operators (SOs), and industry consortia are now specifying the functional requirements and performance criteria for GFMs.%

\textcolor{black}{SOs expect a GFMs to exhibit voltage source characteristics in the (sub)-transient time frame (0–10 cycles)}. This leads to the widely accepted definition of GFMs to behave as a nearly constant internal voltage source (IVS) behind an effective internal impedance by maintaining a constant or nearly constant internal voltage phasor in the sub-transient to transient time frame. \textcolor{black}{Table \ref{tab:GFM_definition} summarizes the GFM definitions proposed by major organizations worldwide. Although there is broad agreement among regulators, SOs and industry consortia, slight variations exist in the terminology used to describe GFM characteristics. The North American Electric Reliability Corporation (NERC) \cite{nerc2021grid}, the Finnish Transmission System Operator (FINGRID) \cite{fingrid2023study}, the Australian Energy Market Operator (AEMO) \cite{aemo2023voluntary}, Universal Interoperability for Grid-Forming Inverters (UNIFI) \cite{unifi2024specs}, and Midcontinent Independent System Operator (MISO) \cite{miso2024gridforming} specify that a GFM should maintain a \enquote{nearly} constant IVS in the (sub)-transient time frame. The Agency for the Cooperation of Energy Regulators (ACER) \cite{acer2023recommendation} also explicitly states that GFMs should behave as a Thevenin source (a voltage source behind an internal impedance). However, the National Energy System Operator (NESO) \cite{eso2021gc0137} specifies that the active power output of a GFM should respond to voltage phase variations at the grid entry point, which also reflects the characteristics of a Thevenin source}. To comply with this requirement, the GFM must exchange the appropriate active and reactive power. For instance, assuming a constant IVS, ENTSO-E proposes compliance testing by generalizing the peak power requirement in response to a voltage change as a function of the effective impedance between the IVS and the point of interconnection (POI). To assess the sensitivity of active and reactive power exchange in case of a voltage phase and magnitude variation at the POI, ENTSO-E recommends a minimum, default, and maximum value of effective impedance \cite{entsoe2024report}. Similar to ENTSO-E, NESO in the UK also requires the active power output of a GFM to be dependent on this effective impedance \cite{eso2021gc0137}. Thus, an accurate measurement and estimation of effective impedance is essential to assess compliance and characterize the expected behavior of the GFM.%

\begin{figure}[htbp]
    \centering
    \includegraphics[width=1\linewidth]{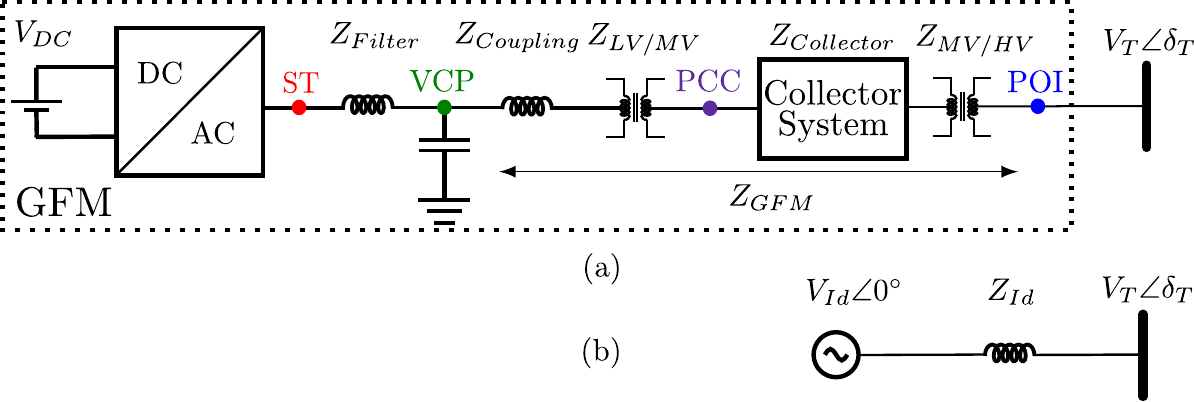}
    \caption{(a) A generic GFM plant connected to POI, (b) Ideal voltage source behind a series impedance ($Z_{Id}$) at POI.}
    \label{fig:GFM_combined}
\end{figure}

Figure \ref{fig:GFM_combined}(a) shows a generic GFM circuit including filter impedance ($Z_{Filter}$), coupling impedance ($Z_{Coupling}$), LV/MV transformer impedance ($Z_{LV/MV}$), collector network impedance ($Z_{Collector}$) and MV/HV transformer impedance ($Z_{MV/HV}$). The GFM plant connects to the network at the point of interconnection (POI). The terminals of the inverter's switch are denoted as ST. The GFM controls regulate the voltage across the filter capacitor, termed as a voltage control point (VCP) - a key point of consideration in defining GFM.%

\textcolor{black}{Although largely consistent, the GFM definitions stated by regulators, SOs, and industry consortia do not explicitly specify the location of the IVS}. To the best of the authors' knowledge, only EPRI \cite{epri2024gridforming} explicitly states that GFM's IVS corresponds to the terminals of individual inverter switches (ST). This suggests the expected characterization of GFM (assuming sufficiently fast voltage control bandwidth and zero virtual impedance) as a nearly constant voltage source ($V_{ST}$) behind a fixed hardware impedance ($Z_{Filter} + Z_{GFM}$) at POI. However, the literature lacks conclusive evidence, and to the best of the author's knowledge, no stability studies have been reported to explicitly substantiate this claim. \textcolor{black}{This lack of clarity and mischaracterization of the location of the IVS and thus the effective impedance can lead to mischaracterization of GFM's expected behavior}. These mischaracterizations of GFM capabilities could lead to unrealistic expectations during compliance, thus risking system reliability. \textcolor{black}{It can also potentially misguide GFM developers and original equipment manufacturers (OEMs), leading to unnecessary costs}.%

\begin{table*}[htbp]
    \centering
    \renewcommand{\arraystretch}{1.5}
    \setlength{\arrayrulewidth}{0.8pt}
    \captionsetup{justification=centering}
    \begin{tabular}{|>{\centering\arraybackslash}m{1.65cm}|
                    >{\centering\arraybackslash}m{0.5cm}|
                    m{14.52cm}|}
        \hline
        \textbf{Organization} & \textbf{Year} & \multicolumn{1}{c|}{\textbf{GFM Definition}} \\
        \hline
        \textbf{NESO \cite{eso2021gc0137}} \newline (GB) & 2021 & A GFM plant's \enquote{\textit{Active Power output is directly proportional to the magnitude and phase of its \textbf{Internal Voltage Source}, the magnitude and phase of the voltage at the Grid Entry Point or User System Entry Point and the sine of the Load Angle.}} \\
        \hline
        \textbf{NERC \cite{nerc2021grid}} \newline (USA) & 2021 & \enquote{\textit{Grid Forming Control for BPS-Connected Inverter-Based Resources are controls with the primary objective of maintaining an \textbf{internal voltage phasor} that is constant or nearly constant in the sub-transient to transient time frame.}} \\
        \hline
        \textbf{FINGRID \cite{fingrid2023study}} \newline (Finland) & 2023 & \enquote{\textit{GFM shall provide autonomous, near-instantaneous frequency and voltage support by maintaining a nearly constant \textbf{internal voltage phasor} in the sub-transient time frame.}} \\
        \hline
        \textbf{AEMO \cite{aemo2023voluntary}} \newline (Australia) & 2023 & \enquote{\textit{A GFM inverter maintains a constant \textbf{internal voltage phasor} in a short time frame, with magnitude and frequency set locally by the inverter, thereby allowing immediate response to a change in the external grid.}} \\
        \hline
        \textbf{ENTSO-E / ACER \cite{acer2023recommendation}} \newline (Europe) & 2023 & \enquote{\textit{Within the power park module’s current and energy limits, the power park module shall be capable of behaving at the terminals of the individual unit(s) as a \textbf{voltage source behind an internal impedance} (Thevenin source), during normal operating conditions (non-disturbed network conditions) and upon inception of a network disturbance.}} \\
        \hline
        \textbf{UNIFI \cite{unifi2024specs}} \newline (USA) & 2024 & \enquote{\textit{GFM IBR controls maintain an \textbf{internal voltage phasor} that is constant or nearly constant in the sub-transient to transient time frame.}} \\
        \hline
        \textbf{MISO \cite{miso2024gridforming}} \newline (USA) & 2024 & \enquote{\textit{GFM IBR shall provide autonomous, near-instantaneous frequency and voltage support by maintaining a nearly constant \textbf{internal voltage phasor} in the sub-transient time frame, within the inverter’s current limits and the resource’s energy limitations.}} \\
        \hline
    \end{tabular}
    \caption{GFM definitions worldwide - Broad agreement to maintain a (nearly) constant IVS in (sub-) transient time frame.}
    \label{tab:GFM_definition}
\end{table*}

This paper addresses this concern by examining the widely available GFM definitions. This is done by comparing the transient response of a GFM to an ideal voltage source (IDVS) under a voltage magnitude and phase jump in the network. The EMT simulations in PSCAD show that a GFM (assuming zero virtual impedance) exhibits the (nearly) constant voltage characteristics across the filter capacitor (VCP) rather than at the terminal of inverter switches (ST). We then present a systematic method for quantifying the effective impedance of the GFM from its black-box model using frequency-domain admittance spectra. This is done by fitting the admittance plot ($Y_{qd}$) of the black-box GFM model to that of an IDVS. \textcolor{black}{Using the standard PSCAD GFM models developed by the National Lab of Rockies (NLR, formerly NREL) \cite{kenyon2021open}, we demonstrate that the obtained GFM's equivalent impedance model accurately represents the active power transfer limits due to static voltage stability and provides a similar sub-transient time-domain response. Further, it is demonstrated that replacing the full GFM model with the proposed equivalent circuit model can accurately capture the small-signal stability characteristics of the modified IEEE-39 bus system \cite{ref39bus}}. %

\section{LOCATION OF INTERNAL VOLTAGE SOURCE}
\subsection{Transient Response of an Ideal Voltage Source}

%
%
Grid disturbances such as faults or sudden changes in generation or load can cause abrupt voltage magnitudes and phase jumps. An IDVS responds by exchanging the required active ($P_{Id}$) and reactive power ($Q_{Id}$) to maintain a constant voltage phasor at its terminals. As shown in Figure \ref{fig:GFM_combined}(b), an IDVS $V_{Id} \angle 0^\circ$ is connected to the grid $V_{T} \angle\delta_T$ via an impedance $Z_{Id} = R_{Id} + jX_{Id}$.  Assuming a step change in grid voltage from $V_T^1 \angle\delta_T^1$ to $V_T^2 \angle\delta_T^2$ at $t=0$, the transient active power response of the IDVS at its terminals is given by \eqref{eq:P_eqn} \cite{PJ_power_German}. Similarly, reactive power response can be derived as \eqref{eq:Q_eqn}. $P_{SS}$ \& $Q_{SS}$ denote the steady-state active and reactive power, respectively, after the transient has settled. The transient power response depends on both the voltage magnitude difference ($V_T^2-V_T^1$) and the phase angle jump ($\delta_T^2-\delta_T^1$) in \eqref{eq:P_eqn} and \eqref{eq:Q_eqn}. In the next sub-section, these analytical equations validate the results from the EMT simulations in PSCAD.%

\begin{figure*}[t]
\begin{linenomath}
\begin{subequations} \label{eq:PQ_eqn}
\begin{align}
P_{\text{Id}}(t) &= P_{\text{SS}}
+ \frac{3}{2} \frac{V_{\text{Id}}}{Z_{Id}} e^{-t/\tau} 
\left[ (V_{T}^{2} - V_{T}^{1}) \cos(\omega t + \phi + \delta_{T}^{1}) 
+ 2 V_{T}^{2} \sin\left(\frac{\delta_{T}^{1} - \delta_{T}^{2}}{2}\right) \sin\left(\omega t + \phi + \frac{\delta_{T}^{1} + \delta_{T}^{2}}{2} \right) \right], \label{eq:P_eqn} \\
Q_{\text{Id}}(t) &= Q_{\text{SS}} + \frac{3}{2} \frac{V_{\text{Id}}}{Z_{Id}} e^{-t/\tau} 
\left[ (V_{T}^{1} - V_{T}^{2}) \sin\left(\omega t + \phi + \delta_{T}^{1} \right) 
+ 2 V_{T}^{2} \sin\left(\frac{\delta_{T}^{1} - \delta_{T}^{2}}{2} \right) \cos\left(\omega t + \phi + \frac{\delta_{T}^{1} + \delta_{T}^{2}}{2} \right) 
 \right]. \label{eq:Q_eqn}
\end{align}
\end{subequations}
\end{linenomath}
\end{figure*}

\subsection{Case Studies}
\begin{figure}[h]
    \centering
    \includegraphics[width=1\linewidth]{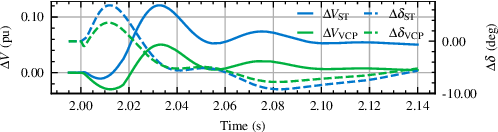}
    \caption{Incremental voltage change at ST and VCP of GFM.}
    \label{fig:GFM_Practical_V_Comp}
\end{figure}

In this section, the GFM’s ability to maintain a constant voltage at its VCP is validated. As shown in Figure \ref{fig:GFM_combined}(a), the collective impedance between VCP and the POI is given as $Z_{GFM} = Z_{Coupling} + Z_{LV/MV} + Z_{Collector} + Z_{MV/HV}$. A simple droop-based GFM is adapted from the NLR GFM model \cite{kenyon2021open}, which regulates voltage at the VCP. In this section, the virtual impedance of the GFM is set to zero. The GFM is tested with a voltage disturbance (both magnitude and phase) at its POI ($V_{T} \angle\delta_{T}$), with $\Delta V_{T} = -0.1 \;pu$ and $\Delta \delta_{T} = -5^\circ$. As shown in Figure \ref{fig:GFM_Practical_V_Comp}, the transient change in $\delta_{VCP}$ is less than $\delta_{ST}$. $V_{ST}$ has a lower variation just after the disturbance; however, at about 30 ms, the peak variation in $V_{ST}$ is more than that $V_{VCP}$. This does not clearly show whether ST or VCP exhibits constant voltage characteristics. To clarify, we compare the transient response of a GFM and an IDVS under similar voltage disturbances. 

Four case studies are presented in Table \ref{tab:case_study_gfm}. The voltage at the network connection point of IDVS is also set at $V_{T} \angle\delta_{T}$. The default typical value for $Z_{Filter}$ is set to 0.15 pu. During the initial steady state, $V_{Id} \angle 0^\circ = V_{VCP} \angle 0^\circ$. The GFM’s current limit is neglected. In Cases I and III, $Z_{GFM}$ is set to 0.2 pu. In Cases I, II, and III, the GFM controls are tuned to make it idealistic, emulating the behavior of an IDVS. This is done by significantly increasing the voltage control bandwidth, disabling the droop control, and ensuring zero virtual impedance. As a result, the idealistic GFM maintains a constant internal frequency and phase angle, regardless of grid disturbances. The response speed of this GFM primarily depends on the voltage control loop bandwidth. The full NLR GFM model in PSCAD is the realistic GFM.%

\begin{table}[htbp]
\renewcommand{\arraystretch}{1.3}
\begin{center}
\caption{Case studies.}
\label{tab:case_study_gfm}
\begin{tabular}{|@{\hskip 2pt}>{\centering\arraybackslash}p{0.5cm}
                |@{\hskip 2pt}>{\centering\arraybackslash}p{1cm}
                |@{\hskip 2pt}>{\centering\arraybackslash}p{1cm}
                |@{\hskip 2pt}>{\centering\arraybackslash}p{5cm}|}
\hline
\textbf{Case} & \textbf{\boldmath$Z_{Id}$}& \textbf{GFM} & \textbf{Variation} \\
\hline
I & Variable & Idealistic & $Z_{Id} = Z_{GFM}$ or $Z_{GFM} + Z_{Filter}$ \\
\hline
II & $Z_{GFM}$ & Idealistic & $0.125 \leq Z_{GFM} \leq 0.333$ \\
\hline
III & $Z_{GFM}$ & Idealistic & $0.075 \leq Z_{Filter} \leq 0.1875$ \\
\hline
IV & $Z_{GFM}$ & Realistic & $0.075 \leq Z_{Filter} \leq 0.1875$ \\
\hline
\end{tabular}
\end{center}
\end{table}
%

\begin{figure}[htbp]
\centering
\begin{minipage}[t]{0.5\linewidth}
    \centering
    \includegraphics[width=\linewidth]{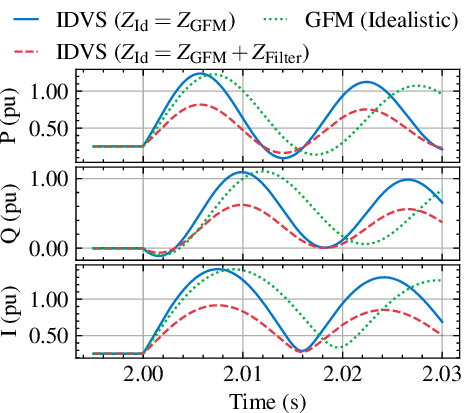}
    \caption*{(a)}
    \label{fig:GFM_Ideal_vs_IVDS_basic}
\end{minipage}%
\hfill
\begin{minipage}[t]{0.5\linewidth}
    \centering
    \includegraphics[width=\linewidth]{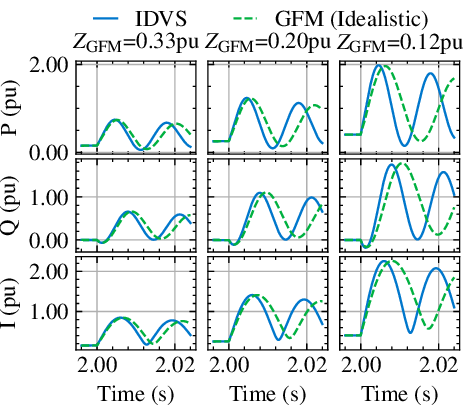}
    \caption*{(b)}
    \label{fig:GFM_Ideal_vs_IDVS_SCR}
\end{minipage}
\caption{(a) Case I - GFM (Idealistic) and IDVS response (b) Case II - varying $Z_{GFM}$: GFM (Idealistic) and IDVS response.}
\label{fig:GFM_IDVS_voltage_and_SCR}
\end{figure}

In Case I, the transient active power, reactive power, and current responses of the idealistic GFM and IDVS with varying $Z_{Id}$ are compared in Figure \ref{fig:GFM_IDVS_voltage_and_SCR}(a). The measurements are taken at VCP and at $V_{Id}$ for the GFM and IDVS, respectively. The EMT simulation results for IDVS are validated using \eqref{eq:P_eqn} and \eqref{eq:Q_eqn}. The idealistic GFM exhibits similar peak active power, peak reactive power, and peak current responses to a voltage disturbance at the POI as compared to the IDVS in case $Z_{Id}$ is set as $Z_{GFM}$. However, the peak power and current responses of the IDVS are significantly lower when $Z_{Id}$ is set to $Z_{GFM} + Z_{Filter}$. The delay in the peak power and current response of GFM is due to the faster voltage control of the IDVS as compared to GFM. \textcolor{black}{This is because an IDVS responds to a voltage disturbance instantaneously without control delay, whereas the response of a GFM is limited by the finite bandwidth of its inner control loops}. Thus, the GFM exhibits the constant voltage characteristic at an electrical distance of $Z_{GFM}$ from POI, i.e., at the VCP rather than at its ST ($Z_{GFM} + Z_{Filter}$ away from POI). However, given the understanding that a GFM exhibits the constant voltage characteristics at ST, ENTSO-E generalizes this peak power response as a function of effective impedance ($Z_{GFM} + Z_{Filter}$) \cite{entsoe2024report}. This may result in relaxed requirements for compliance testing of the voltage source behavior of GFM plants, potentially leading to system instability. %

\begin{figure}[htbp]
    \centering
    \includegraphics[width=1\linewidth]{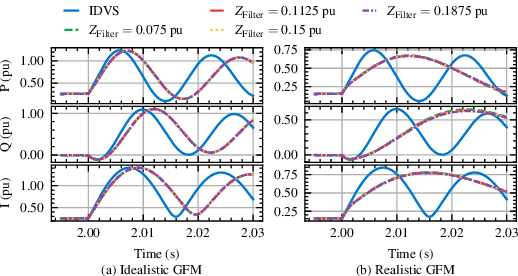}
    \caption{Varying $Z_{Filter}$: (a) Case III - Idealistic GFM, (b) Case IV - Realistic GFM.}
    \label{fig:GFM_vs_IDVS_ZFilter}
\end{figure}

Since different GFM plants may have varying values of $Z_{GFM}$, Case II validates the definition across a broad range of $Z_{GFM}$. Figure \ref{fig:GFM_IDVS_voltage_and_SCR}(b) confirms that, even with varying $Z_{GFM}$, the GFM exhibits the constant voltage characteristic at VCP. Further, in Case III, $Z_{Filter}$ is varied, while the gains are adjusted to maintain 300 Hz current control bandwidth. $Z_{GFM}$ and other control gains are kept constant. Figure \ref{fig:GFM_vs_IDVS_ZFilter}(a) shows that changing $Z_{Filter}$ has virtually no impact on the peak power, current, or time-to-peak response. Case IV further validates this using the realistic (full NLR) GFM model. $Z_{GFM} = Z_{Id}$ is set to 0.33 pu. Results are obtained by varying $Z_{Filter}$ while keeping other parameters constant. As shown in Figure \ref{fig:GFM_vs_IDVS_ZFilter}(b), due to the slower voltage and droop control dynamics of the realistic GFM, its peak power and current responses are lower than those of the IDVS. However, they remain consistent across a wide range of $Z_{Filter}$. Similarly, there is no notable change in time-to-peak response. These results confirm the hypothesis that a GFM maintains a (nearly) constant voltage phasor at the VCP rather than at its ST.

\begin{figure}[htbp]
    \centering
    \includegraphics[width=1\linewidth]{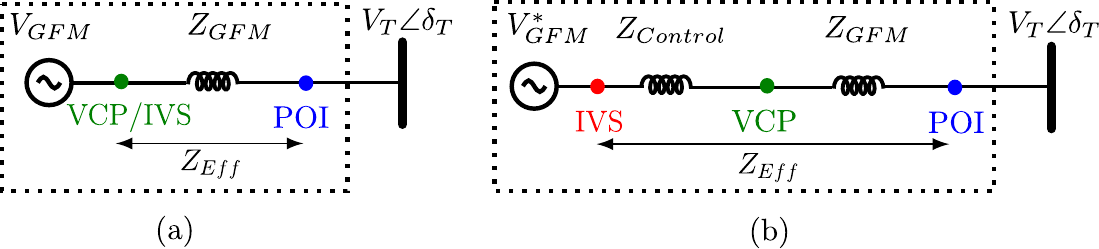}
    \caption{Equivalent GFM plant (a) without virtual impedance, (b) with virtual $Z_{Control}$.}
    \label{fig:GFM_plant_eq}
\end{figure}

Therefore, a GFM (without any virtual impedance) can be represented as a (nearly) constant voltage source $V_{GFM}$ behind effective impedance $Z_{GFM}$, as shown in Figure \ref{fig:GFM_plant_eq}(a). This resembles synchronous machine model of an internal voltage source ($E$) behind an impedance ($X$); however $E$ and $X$ vary after a disturbance and are classified in steady state ($E_S$, $X_S$), transient ($E^{\prime}$, $X^{\prime}$) or sub-transient ($E^{\prime\prime}$, $X^{\prime\prime}$) time frame. More holistically, considering the equivalent virtual impedance, the GFM representation can be refined to a constant voltage source $V_{GFM}^*$ behind the combined effective impedance $Z_{Control} + Z_{GFM}$ \cite{impedace_circuit_modelling}, as illustrated in Figure \ref{fig:GFM_plant_eq}(b). In the next section, we show how the equivalent impedance of a GFM at its POI can be obtained from the GFM black box model. 

\section{Equivalent Impedance Model of the GFM}

ENTSO-E has proposed detailed technical requirements for evaluating GFM compliance at the POI. Regarding the voltage source behavior of a GFM, it specifies that a GFM shall be "capable of behaving at the terminals of the individual unit(s) as a voltage source behind an internal impedance (Thevenin source)" \cite{entsoe2024report}. To ensure the generality of the power requirement in case of a grid disturbance, ENTSO-E defines it as a function of the effective reactance ($X_{Eff}$). i.e, the total reactance between the IVS and the POI. $X_{Eff}$ accounts for total virtual impedance, which could be explicitly designed, equivalent control-based impedance, or both. This representation is equivalent to modeling a constant voltage source $V_{GFM}^*$ behind the combined impedance $Z_{Control} + Z_{GFM} = Z_{Eff}$, as illustrated in Figure \ref{fig:GFM_plant_eq}(b).

The variation in output current/power of the GFM during a voltage angle or magnitude jump at the POI depends on $X_{Eff}$. To assess the expected GFM response, ENTSO-E has proposed minimum, default, and maximum values of $X_{Eff}$ at 50 Hz \cite{entsoe2024report}. Table \ref{tab:xeff_values} lists the recommended values for different connection points. \textcolor{black}{The $X_{Eff} / R_{Eff}$ ratio of 10 is recommended}. However, since vendor-specific GFM models are mostly black-box representations, the grid operators need a systematic method to evaluate the equivalent $Z_{Eff}$ of a GFM.

\begin{table}[ht]
\centering
\caption{ENTSO-E's proposed range of $X_{Eff}$ (at 50 Hz) for compliance at different locations of the POI \cite{entsoe2024report}.}
\label{tab:xeff_values}
\renewcommand{\arraystretch}{1.2}
\begin{tabularx}{\linewidth}{>{\centering\arraybackslash}X c c c}
\toprule
\textbf{Location of POI} 
& \textbf{Min (p.u.)} 
& \textbf{Default (p.u.)} 
& \textbf{Max (p.u.)} \\ 
\midrule
LV terminals & $0.17$ & $0.25$ & $0.27$ \\ 
MV terminals & $0.25$ & $0.33$ & $0.35$ \\ 
HV terminals & $0.40$ & $0.48$ & $0.50$ \\ 
\bottomrule
\end{tabularx}
\end{table}

\subsection{Admittance Spectra}

For any resource, the frequency-domain relationship between the $dq$ components of voltage and current at the POI is given in (\ref{eq:admittance}). The negative sign in (\ref{eq:admittance}) indicates the source convention at the POI. 

\begin{equation}
- \begin{bmatrix}
I_d(s) \\
I_q(s)
\end{bmatrix}
=
\begin{bmatrix}
Y_{dd}(s) & Y_{dq}(s) \\
Y_{qd}(s) & Y_{qq}(s)
\end{bmatrix}
\begin{bmatrix}
V_d(s) \\
V_q(s)
\end{bmatrix}
\label{eq:admittance}
\end{equation}

As shown in \cite{impedance_modelling_shahil}, for small-signal analysis, a perturbation in voltage magnitude is equivalent to a perturbation in the corresponding $d$-axis voltage, i.e., $V_{m}(s) = V_{d}(s)$. Thus, the term $Y_{qd}$ represents the reactive power response of the device to a voltage magnitude perturbation at the POI. Similarly, the term $Y_{dq}$ represents the active power response to a voltage phase perturbation at the POI. The admittance matrix for an IDVS with $Z_{Id} = R_{Id}+sL_{Id}$ is given in (\ref{eq:Y_IDVS}) \cite{testing_shahil_shah}. 

\begin{figure}[htbp]
    \centering
    \includegraphics[width=0.9\linewidth]{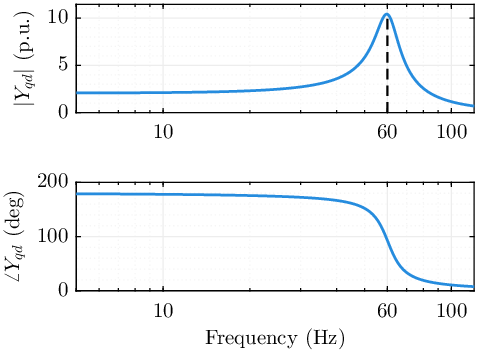}
    \caption{$Y_{qd}$ admittance plot for IDVS with proposed default values at HV terminals: $X_{Eff} = 0.48$ p.u., $X_{Eff}/R_{Eff} = 10$}
    \label{fig:bode_Ydq_thevenin}
\end{figure}

\begingroup
\small
\setlength{\arraycolsep}{1pt}
\renewcommand{\arraystretch}{0.9}
\begin{equation}
Y_{Id}(s)=\frac{1}{(R_{Id}+sL_{Id})^{2}+(\omega_1L_{Id})^{2}}
\begin{bmatrix}
R_{Id}+sL_{Id} & \omega_1L_{Id}\\
-\omega_1L_{Id} & R_{Id}+sL_{Id}
\end{bmatrix}
\label{eq:Y_IDVS}
\end{equation}
\endgroup

where, $\omega_1$ is the fundamental frequency. The term $Y_{qd}$, is a second-order transfer function with poles at $s = -\tfrac{R_{Id}}{L_{Id}} \pm j \omega_{1}$ and a damped natural frequency of $\omega_d = \omega_1$. This appears as a resonance peak at fundamental frequency in the Bode plot of $Y_{qd}$ in Figure \ref{fig:bode_Ydq_thevenin}. The steady-state (low-frequency) response has a magnitude of $|Y_{qd}(0)| = \dfrac{\omega_{1} L_{Id}}{R_{Id}^{2} + (\omega_{1} L_{Id})^{2}}$. A phase of $180^\circ$ at low frequencies indicates that the IDVS exchanges reactive power to oppose changes in voltage magnitude at POI. 

\subsection{Curve Fitting}
For a GFM to behave like a Thevenin source as specified by ENTSO-E, it must exhibit similar admittance characteristics in the frequency range of interest. To achieve comparable time-domain behavior in the (sub)-transient time frame (0–10 cycles), the frequency range of interest is approximately 5-100 Hz. The lower bound of 5 Hz is derived from the end of the (sub)-transient interval. Frequencies above 100 Hz are neglected, since current control loops and network dynamics act much faster and do not influence the (sub)-transient response. Also, since the synchronization and droop control loops are slower, the admittance transfer functions remain operating-point independent within this frequency range \cite{impedace_circuit_modelling}.

To obtain the equivalent impedance model of the GFM, a curve-fitting approach is used over the frequency range of 5–100 Hz. First, admittance spectra are generated from the black-box GFM models in PSCAD by modulating the $d$- and $q$-axis voltages at fixed frequencies. The steady-state time-domain response is then post-processed using the Fast Fourier Transform (FFT) to extract the magnitude and phase shift at each perturbation frequency. A Python-based toolbox, Z-Tool \cite{ztool}, is used to compute the frequency-domain admittances of the black-box device at the POI. The measured $Y_{qd}$ admittance of the full model of the device at each perturbation frequency ($f_k$) is denoted as $Y_{qd}^{full}(f_k)$. At the same perturbation frequencies, the $Y_{qd}$ admittance of the Thevenin equivalent is given in (\ref{eq:Y_qd_eq}).

\begin{equation}
Y_{qd}^{th}(f_k;R_{th},L_{th}) =
\frac{-\,\omega_1 L_{th}}{(R_{th} + j2\pi f_k L_{th})^2 + (\omega_1 L_{th})^2}
\label{eq:Y_qd_eq}
\end{equation}

The objective is to determine the values of $R_{th}$ and $L_{th}$ such that the root-mean-square (RMS) error $\mathcal{J}(R_{th},L_{th})$ between $Y_{qd}^{th}(f_k;R_{th},L_{th})$ and $Y_{qd}^{full}(f_k)$ is minimized. The corresponding minimization function is given in (\ref{eq:obj_min}).

\begin{multline}
\min_{R_{th},\,L_{th}} \; \mathcal{J}(R_{th},L_{th}) = \\
\min\sqrt{\sum_{f_{\min}}^{f_{\max}}\frac{1}{2\pi f_k}
\left(\frac{\big|Y_{qd}^{th}(f_k;R_{th},L_{th})\big|
- \big|Y_{qd}^{full}(f_k)\big|}
{\big|Y_{qd}^{full}(f_k)\big|}\right)^{\!2}}
\label{eq:obj_min}
\end{multline}


The step response of GFMs in the sub-transient time frame is influenced by the design of the voltage-control loop and virtual impedance, and is dominated by lower frequency dynamics. To ensure that the effect of control design is accurately reflected in the fitted admittance curves, the error function is weighted such that lower frequencies are emphasized, i.e., the RMS error at each perturbation frequency is divided by $2\pi f_k$. Furthermore, the RMS error is minimized between the magnitude $\left| Y_{qd} \right|$ (in pu) rather than the Bode gain $\log_{10}\!\left|Y_{qd}\right|$ (in dB). This choice avoids compressing larger admittance values, since the logarithmic operator $\log_{10}$ relatively amplifies values near zero. Such amplification would increase error sensitivity and result in a poorer fit for devices with higher admittance.


\begin{equation}
(R_{Eff},L_{Eff}) = \arg\min_{R_{th},\,L_{th}} \mathcal{J}(R_{th},L_{th})
\label{eq:argmin}
\end{equation}


The RMS error $\mathcal{J}(R_{th}, L_{th})$ is minimized using the \textit{fmincon} algorithm in MATLAB. As shown in (\ref{eq:argmin}), the resulting $(R_{Eff}, L_{Eff})$ are the equivalent Thevenin or effective impedance values. The maximum error threshold $\epsilon$ is a tunable parameter under the control of SOs that determines the goodness of fit of the device characteristics relative to the obtained Thevenin equivalent. If the values of $R_{Eff}$ and $L_{Eff}$ fall outside the constrained limits $[L_{Eff}^{min}, L_{Eff}^{max}]$ given in Table \ref{tab:xeff_values}, this indicates that, for the specified error threshold $\epsilon$, the GFM does not satisfy the effective impedance compliance criteria.








\subsection{Case Studies \& Results}
In this section, we first demonstrate the validity of the proposed method for obtaining the equivalent impedance model of an IDVS and a synchronous machine. We then show that the GFM equivalent impedance model accurately reproduces the sub-transient reactive power response to a voltage magnitude jump at the POI. \textcolor{black}{This equivalent model can provide insights into the voltage stability of IBR-dominated power systems}. This case study focuses on steady-state analysis to evaluate voltage-stability boundaries associated with active power transfer, represented by well-understood $P$–$V$ curves \cite{PV_Curve}. The declining profile of a $P$–$V$ curve demonstrates that as active power transfer increases, reactive power losses intensify, causing the voltage to decrease. The resulting increase in both active and reactive currents, driven by higher reactive losses, ultimately determines the maximum active power transfer limit \cite{glover2008power}. The proposed GFM equivalent model is shown to accurately capture the active power transfer limits imposed by static voltage stability. \textcolor{black}{Similarly, the equivalent modeling of the GFM can serve as an effective screening tool to provide quick insights into potential small-signal stability issues and identify a limited set of vulnerable cases, without requiring time-consuming and computationally intensive EMT simulations. To illustrate this, it is shown that replacing the full GFM model with the proposed equivalent circuit model can reproduce the small-signal stability characteristics of the modified IEEE-39 bus system \cite{ref39bus} with reasonable accuracy. However, it should be noted that detailed studies for this limited set of cases should be performed using full EMT models of the GFM. It is also worth noting that the equivalent circuit model of the GFM capturing its sub-transient behavior is derived from detailed EMT models of the device and therefore inherently captures the dynamics represented in EMT simulations. Consequently, the proposed model also encompasses the sub-transient dynamics typically represented in traditional RMS models}. 

\subsubsection{Ideal Voltage Source \& Synchronous Machine}

\begin{figure}[htbp]
    \centering
    \includegraphics[width=0.9\linewidth]{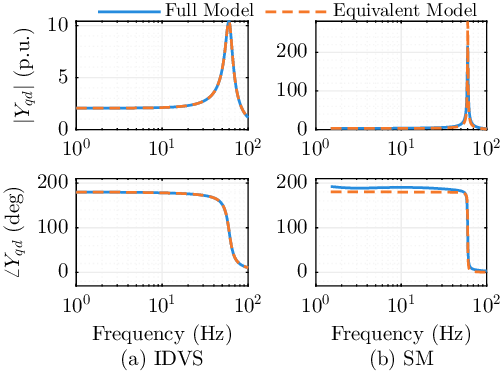}
    \caption{$Y_{qd}$ admittance spectra for full vs equivalent models of (a). an IDVS and (b) a SM}
    \label{fig: idvs_sm_fitted}
\end{figure}

The admittance spectra for a black-box device (an IDVS and an SM in this case) are obtained in PSCAD using the Python-based Z-Tool toolbox \cite{ztool}. The IDVS has a reactance of 0.48 p.u. with an $X/R$ ratio of 10. The $Y_{dq}$ admittance spectra from the full and equivalent model are shown in Figure \ref{fig: idvs_sm_fitted}(a). The perfect fit, with an RMS error of 0.00\%, yields $R_{Eff}^{IDVS} = 0.048$ p.u. and $X_{Eff}^{IDVS} = 0.48$ p.u.. Figure \ref{fig: idvs_sm_fitted}(b) shows $Y_{dq}$ admittance spectra for the full and equivalent model of an SM, measured at the SM terminals. The SM model used in PSCAD is a sixth-order 200 MVA machine with a DC1A excitation system and a mechanical-hydraulic governor. The SM model has sub-transient reactance $x_d^{''}$ $= 0.25$ p.u. and armature resistance $r_a$ $= 0.0025$ p.u.. The $Y_{dq}$ admittance spectra of the SM exhibits a constant response at low frequencies with a phase of $-180^\circ$, similar to the IDVS $Y_{dq}$ admittance. Curve fitting with an RMS error of 0.27\% gives $R_{Eff}^{SM} = 0.0024$ p.u. and $X_{Eff}^{SM} = 0.266$ p.u.. This validates the accuracy of the proposed method for obtaining the equivalent circuit of a device in the sub-transient time frame. Therefore, the method can also be used to determine the sub-transient reactance of an SM, eliminating the need for traditional short-circuit tests.
\subsubsection{GFM}

In this subsection, the equivalent impedance model of a GFM model developed in PSCAD by NLR is obtained. The control parameters of this model, described in \cite{kenyon2021open}, are adjusted within the usual bandwidths \cite{chatterjee2025unifi}. However, these models are treated as black-box representations at their POI. The combined collector system impedance $Z_{Collector}$ and the MV/HV transformer impedance $Z_{MV/HV}$ together equals 0.30 p.u., with an $X/R$ ratio of 10. The admittance is measured at the POI.

\begin{figure}[htbp]
    \centering
    \includegraphics[width=0.9\linewidth]{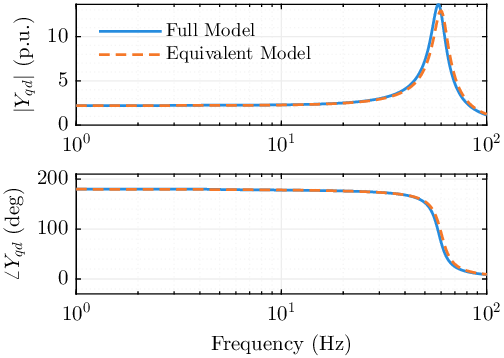}
    \caption{$Y_{qd}$ admittance spectra for full vs equivalent models of NLR GFM Model in PSCAD}
    \label{fig:NREL_Full_SCR_0_3}
\end{figure}


$Y_{qd}$ admittance spectra for full vs equivalent models of the NLR GFM model are shown in Figure \ref{fig:NREL_Full_SCR_0_3}. Similar to the IDVS and SM, the GFM exhibits a flat magnitude $|Y_{dq}|$ at low frequencies. The near $180^\circ$ phase at these frequencies indicates that the GFM opposes voltage changes at the POI by exchanging reactive power. However, the resonance peak of the GFM $Y_{dq}$ admittance occurs at 58.01 Hz, lower than the 60 Hz observed for the IDVS and SM. The response for the equivalent NLR GFM model yields an RMS error of 0.037\%, with $R_{Eff}^{GFM} = 0.038$ p.u. and $X_{Eff}^{GFM} = 0.447$ p.u.. This GFM satisfies the ENTSO-E effective impedance compliance criteria, as the value $X_{Eff}^{GFM} = 0.447$ p.u. lies within the range [0.40, 0.50] at the HV terminals. The NLR GFM model also has an equivalent $X/R$ ratio of 11.37, which is close to the ENTSO-E assumption of 10.

\subsubsection{\textcolor{black}{Impact of GFM Design on Equivalent Circuit Model}}
\begin{figure}[htbp]
    \centering
    \includegraphics[width=0.9\linewidth]{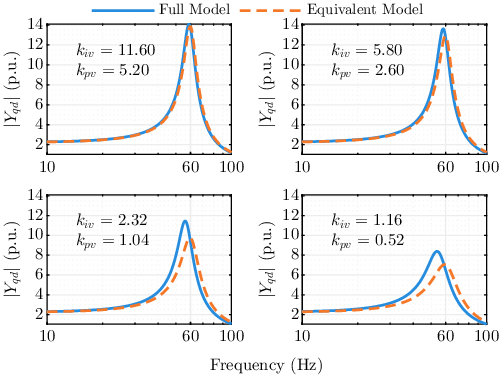}
    \caption{$Y_{qd}$ admittance spectra for full vs equivalent models of GFM with changing voltage control gains}
    \label{fig:VBW_Compare}
\end{figure}

The shape of $Y_{dq}$ admittance curves and their resonance frequency within the 5–100 Hz range depends on the control design and the voltage-loop control gains of the GFM \cite{epri2024gridforming}. The synchronization and droop control loops are generally slower and therefore do not affect the admittance characteristics in this frequency range of interest \cite{epri2024gridforming}. Table \ref{tab:v_band_compare} shows the variation in the equivalent impedance of the NLR GFM with changes in voltage-control loop gains. The proportional gain $k_{pv}$ and integral gain $k_{iv}$ are varied in a fixed ratio. The voltage-control bandwidth directly influences the voltage-source strength of the GFM in the sub-transient time frame, and thus is reflected in its equivalent impedance. As shown in Figure \ref{fig:VBW_Compare}, reducing the voltage-control loop gains shifts the resonance frequency of $Y_{dq}^{GFM}$ downward, moving it away from 60 Hz. This shift increases the RMS error, since an IDVS (and the Thevenin equivalent of a GFM) always has its resonance peak at 60 Hz. Thus, a higher RMS error in this case also indicates weaker voltage-source characteristics of the GFM. With decreasing voltage-control bandwidth, the equivalent impedance of the GFM increases, with the most significant change observed in $R_{Eff}^{GFM}$. This behavior is consistent with the analytically derived equivalent circuit model for droop-based GFMs in \cite{impedace_circuit_modelling}.

\begin{table}[htbp]
\centering
\caption{Effect of changing voltage control loop gains on effective impedance (in p.u.) of the GFM}
\label{tab:v_band_compare}
\begin{tabular}{c c c c c}
\hline
$\{k_{iv},\,k_{pv}\}$(p.u.) 
& $\omega_{res}$ (Hz)
& $X_{Eff}$ 
& $R_{Eff}$ 
& RMS Error\\ 
\hline
$\{11.60,\, 5.20\}$ & 59.06 & 0.44 & 0.03 & 0.03\% \\ 
$\{5.80,\, 2.60\}$   & 58.01 & 0.44 & 0.03 & 0.03\% \\ 
$\{2.32,\, 1.04\}$ & 56.06 & 0.44 & 0.05 & 0.06\% \\ 
$\{1.16,\, 0.52\}$ & 54.02 & 0.43 & 0.07 & 0.08\% \\ 
\hline
\end{tabular}
\end{table}

\subsubsection{\textcolor{black}{Validation of Equivalent Circuit Model}}
\begin{figure}[htbp]
    \centering
    \includegraphics[width=0.9\linewidth]{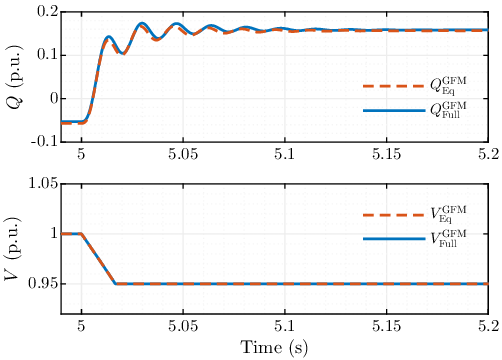}
    \caption{Reactive power response of full GFM model vs Thevenin equivalent to a 5\% voltage change at its POI}
    \label{fig:Q_response_V}
\end{figure}

To validate the accuracy of the obtained GFM equivalent impedance model, Figure \ref{fig:Q_response_V} compares the reactive power response of the equivalent model with that of the full NLR GFM model following a voltage-magnitude step change at the POI. In steady state, both the GFM and its Thevenin equivalent supply 0.4 p.u. of active power and absorb 0.05 p.u. of reactive power at the POI. At $t = 5$ seconds, the voltage magnitude at the POI of the full GFM model ($V_{Full}^{GFM}$) and the Thevenin equivalent ($V_{Eq}^{GFM}$) is reduced by 5\% from 1.0 p.u. to 0.95 p.u. Holding its IVS voltage constant, the GFM responds by increasing its reactive power output. Since the IVS voltage cannot be measured in a black-box model, ENTSO-E requires that compliance be demonstrated at the POI. As expected, the reactive power responses of the full GFM model ($Q_{Full}^{GFM}$) and the Thevenin equivalent ($Q_{Eq}^{GFM}$) are very similar. In the (sub)-transient time frame (first ~200 ms after the disturbance), the RMS error between $Q_{Eq}^{GFM}$ and $Q_{Full}^{GFM}$ is 2.76\%.


\subsubsection{\textcolor{black}{Application I - Static Voltage Stability Limit }}
\begin{figure}[htbp]
    \centering
    \includegraphics[width=0.9\linewidth]{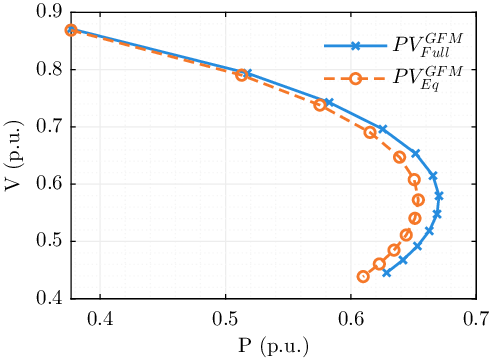}
    \caption{PV curve for full GFM model vs Thevenin equivalent}
    \label{fig:PV_Curve}
\end{figure}

To demonstrate the application of the equivalent GFM model, this subsection compares the voltage-stability boundary associated with active power transfer. The 200 MW NLR GFM (and its Thevenin equivalent) is connected to a constant-power load. At the initial steady-state operating point, the active power ($P_{Load}$) and reactive power ($Q_{Load}$) base loads are 25 MW and 12.5 MVAr, respectively. To test the active power transfer limit, the active and reactive loads are increased in steps of 10 MW and 5 MVAr, respectively. The resulting $P$–$V$ curves for the full GFM model ($PV^{GFM}_{Full}$) and the Thevenin equivalent model ($PV^{GFM}_{Eq}$) are shown in Figure \ref{fig:PV_Curve}. The two $P$–$V$ curves are very similar; however, the difference between the two traces is exaggerated by the zoomed x-axis. The maximum active power transfer capability of the full GFM model ($P^{Max}_{Full}$) is 0.670 p.u., occurring at a voltage of $V^{P{Max}}_{Full} = 0.58$ p.u. Similarly, the equivalent GFM model reaches its maximum active power transfer capability of $P^{Max}_{Eq} = 0.653$ p.u. at $V^{P_{Max}}_{Eq} = 0.57$ p.u.. The error in active power transfer capability is only 2.46\%. The equivalent GFM model does not guarantee a strictly conservative estimate; however, it provides a reliable and accurate assessment of static voltage stability for power system planners. 

\subsubsection{\textcolor{black}{Application II - Small Signal Stability}}

\begin{figure}[htbp]
    \centering
    \includegraphics[width=1\linewidth]{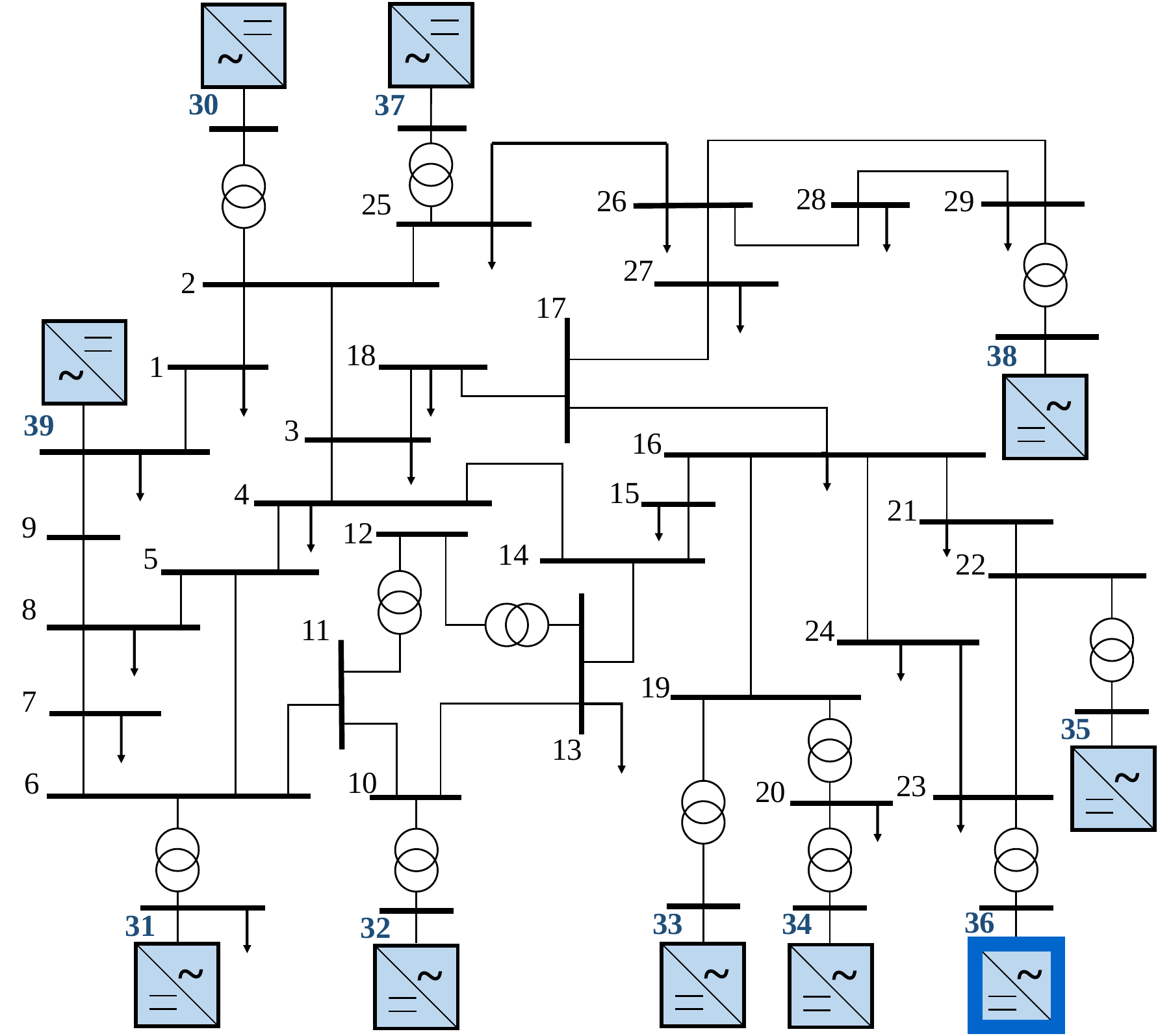}
    \caption{\textcolor{black}{A modified IEEE 39-bus test system \cite{ref39bus}, with all machines replaced by IBRs and a GFM at Bus 36.}}
    \label{fig:39Bus_c}
\end{figure}

\textcolor{black}{Here we show that replacing the full GFM model with the proposed equivalent circuit model can reproduce the small-signal stability characteristics of the modified IEEE 39-bus system \cite{ref39bus} with reasonable accuracy. The IEEE 39-bus system is modified by replacing all 10 synchronous generators with IBRs, as shown in Fig.~\ref{fig:39Bus_c}. To highlight the impact of GFM equivalent modeling on small-signal stability, the 39-bus system is modified to include only one GFMI at Bus~36, while all other generators are replaced by grid-following inverters (GFLs). The IBR configurations and dispatch scenario for buses 30--39 are summarized in Table~\ref{tab:gfl_gfm_scenarios}. The GFL and GFM models used in this case study are adopted from \cite{emt-rms2}. It should be noted that fully detailed (white-box) GFM models are often unavailable to SOs and are not required for equivalent modeling using the proposed method. In this work, these models are treated as black-box representations for equivalent-circuit modeling, while detailed white-box models are used only for system-level modal analysis to provide a ground-truth benchmark}.

\begin{table}[htbp]
\centering
\caption{\textcolor{black}{IBR configurations and dispatch scenario at IBR buses 30--39 in the modified IEEE 39-bus system.}}
\label{tab:gfl_gfm_scenarios}
\renewcommand{\arraystretch}{1.15}
\setlength{\tabcolsep}{3pt}
\footnotesize
\resizebox{\columnwidth}{!}{%
\begin{tabular}{|l|c|c|c|c|c|c|c|c|c|c|}
\hline
\textbf{Bus}   & \textbf{30}  & \textbf{31}  & \textbf{32}  & \textbf{33 } & \textbf{34}  & \textbf{35}  & \textbf{36}  & \textbf{37}  & \textbf{38} & \textbf{39}  \\ \hline
\textbf{Type}  & GFL & GFL & GFL & GFL & GFL & GFL & \textit{GFM} & GFL & GFL & GFL \\ \hline
\textbf{P (pu)} & 2.75 & 5.21 & 6.50 & 6.32 & 5.08 & 6.50 & 5.60 & 5.40 & 8.30 & 10.29 \\ \hline
\textbf{Q (pu)} & 1.88 & 1.57 & 1.69 & 0.94 & 1.59 & 4.72 & 2.73 & 0.32 & 0.64 & 3.56 \\ \hline
\end{tabular}%
}
\end{table}

\textcolor{black}{To assess the impact of replacing the full GFM model with its equivalent on small-signal stability, two cases are considered. In Case A, Bus~36 is equipped with the full GFM model, while in Case B, Bus~36 is equipped with the equivalent impedance-circuit model of the GFM. The equivalent impedance-circuit model of the GFM at Bus~36 is obtained following the procedure described in Section~III. The equivalent modeling yields an RMS error of 0.047\%, resulting in $R_{Eff}^{GFM} = 0.016$~p.u. and $X_{Eff}^{GFM} = 0.149$~p.u.}.

\begin{figure}[htbp]
    \centering
    \includegraphics[width=0.9\linewidth]{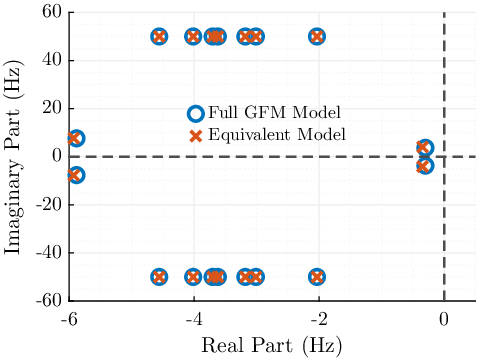}
    \caption{\textcolor{black}{Critical modes of the modified IEEE 39-bus system for Case A (with full GFM model) and Case B (with equivalent GFM model).}}
    \label{fig:eigen}
\end{figure}

\textcolor{black}{System-level modal analysis of Case~A (with the full GFM at Bus~36) and Case~B (with the equivalent model at Bus~36) is presented next. Fig.~\ref{fig:eigen} illustrates the critical oscillatory modes for both cases. It can be observed that Case~B, with the equivalent GFM model, captures all the critical oscillatory modes with reasonable accuracy, including both sub-synchronous and network modes. Similarly, Table~\ref{tab:mode_comparison} compares the frequency and damping factor of the two most critical sub-synchronous oscillatory modes of the system. The critical $\sim$4~Hz mode and the well-damped $\sim$7.7~Hz mode are in close agreement. As described in Section~III, the curve fitting is performed over the frequency range of 5-100~Hz to obtain an equivalent GFM model valid within sub-transient time frames. Therefore, the equivalent model is not expected to capture very slow (low-frequency) modes of the GFM, which are mainly driven by slower outer synchronization control loops of the GFM}. 

\textcolor{black}{It can also be observed from Table \ref{tab:mode_comparison} that the $\sim$4 Hz mode has better damping for case B, while the $\sim$7.7 Hz mode has better damping for Case A. This indicates that the equivalent GFM model does not guarantee a strictly conservative estimate; however, it provides a reliable and accurate initial assessment and can serve as an effective screening tool for small-signal stability for power-system planners. The tool can identify a limited set of vulnerable cases without requiring time-consuming and computationally intensive EMT simulations. However, it should be noted that detailed studies for this limited set of cases should be performed using full EMT models of the GFM}.

\begin{table}[htbp]
\centering
\caption{\textcolor{black}{Comparison of critical oscillatory modes for Case A (with full GFM model) and Case B (with equivalent GFM model).}}
\label{tab:mode_comparison}
\scriptsize
\setlength{\tabcolsep}{5.3pt}
\begin{tabular}{|c|c|c|c|c|c|}
\hline
\multicolumn{2}{|c|}{\textbf{Mode - Eigen Value (Hz)}} & \multicolumn{2}{c|}{\textbf{Frequency}} & \multicolumn{2}{c|}{\textbf{Damping Ratio}} \\
\cline{1-6}
Case A & Case B & Case A & Case B & Case A & Case B \\
\hline
$-0.30 \pm 3.72i$ & $-0.35 \pm 4.08i$ & 3.71 Hz & 4.07 Hz & 8.07\% & 8.47\% \\
\hline
$-5.88 \pm 7.64i$ & $-5.93 \pm 7.71i$ & 7.63 Hz& 7.70 HZ & 61.03\% & 60.96\% \\
\hline
\end{tabular}
\end{table}

\section{Conclusion}

Revisiting the “voltage source behind an impedance” characterization of a GFM, this paper confirms that an idealistic GFM maintains a nearly constant voltage at the voltage control point (VCP), instead of the inverter switch terminals (ST). This distinction is shown to be important when analyzing a GFM's response to grid disturbances. If unaddressed, it could mislead GFM developers and manufacturers to unrealistic expectations and costs, and may also compromise system stability and reliability. Further, it is demonstrated that the effective impedance of a GFM can be estimated from its black-box model using frequency-domain admittance spectra at the POI. \textcolor{black}{The resulting equivalent impedance model can accurately capture the GFM’s sub-transient behavior, static voltage stability limit, and small-signal stability issues. The admittance-based approach can be used to determine the sub-transient reactance of an SM, eliminating the need for traditional short-circuit tests. This approach provides system operators with a reliable and accurate initial assessment, and can serve as an effective screening tool to quickly assess the dynamic interaction between GFMs and the grid without relying on computationally intensive EMT simulations}.

\vspace{-5pt}

\bibliography{GFM_Def_References}




\end{document}